\documentclass[amsmath,showpacs,prl]{revtex4}
\begin{document}

\title{The scaling of conductance in the Anderson model of localization in one dimension is a two-parameter scaling}
\author{Jean Heinrichs}
\email{J.Heinrichs@ulg.ac.be}
\affiliation{Institut de Physique, B{5}, Universit\'{e} de Li\`{e}ge, Sart
Tilman, B-{4000} Li\`{e}ge, Belgium}

\begin{abstract}
The cumulants of the logarithm of the conductance (lng) in the localized regime in the one-dimensional Anderson model are
calculated exactly in the second Born approximation for weak disorder.  Only the first two cumulants turn out to ne non-zero
since the third and fourth cumulants vanish identically and the higher cumulants are of higher order in the disorder.  The
variance and the mean of lng vary linearly with length $L$ while their ratio is proportional to the inverse localization
length. The resulting exact log-normal distribution of conductance thus corresponds to a special form of two-parameter
scaling.  This contradicts the standard one-parameter scaling in the random phase approximation.

\end{abstract}


\maketitle

The appearance of the scaling theory of localization in $d$-dimensional disordered systems \cite{1} and more detailed
developments of it in 1D \cite{2,3,4,5} and for quasi 1D systems \cite{6,7} has inaugurated a golden age for mesoscopic
physics.  This is demonstrated by the numerous review articles \cite{8,9,10,11,12,13} and monographs \cite{14,15} dealing with
application of scaling ideas to transport in disordered conductors.

The fundamental hypothesis in the scaling theory \cite{1} is that the scaling of the logarithm of a typical conductance $g$ as
a function of a characteristic size $L$ of the system is described asymptotically for large $L$ by a universal function,
$\beta(\ln g)$, of a single parameter (SPS), namely $\ln g$ itself:

\begin{equation}\label{eq1}
\frac{d\ln g}{d\ln L}=\beta (\ln g)\quad .
\end{equation}
The function $\beta$ which may generally depend on dimensionality is independent of $L$ and of any microscopic parameter in
the system.  We recall that in the studies of scaling in 1D systems \cite{2,3,4,5} the parameter $\ln ((1+\rho)$, with
$\rho=\frac{1}{g}$ the resistance, was identified as the convenient scaling variable both in the low resistance ($\rho<<1$)
(quasi-metallic) regime and in the large resistance or insulating regime ($\rho >>1$).  This variable reduces to $-\ln g$ in
Eq. (\ref{eq1}) for $\rho>>1$ (localized regime) and, thanks to the Landauer formula, $\rho=\frac{r_L}{t_L}$ (with $r_L$ and
$t_L$ the reflection and transmission coefficients od the system, respectively), it coincides with $-\ln t_L\simeq -\ln g$.

In Refs. \cite{2,3,4,5} on 1D systems it was argued that the scaling theory of Abrahams et al. \cite{1} had to be interpreted
in terms of the scaling of the distribution $P_g(g)$ of the random conductance of the system.  SPS then means that $P_g(g)$ is
fully determined by a single parameter such as e.g. the mean logarithm, $\langle-\ln g \rangle$, which is itself defined by a
scaling equation of the form (\ref{eq1}) with $\ln g$ replaced by $\langle\ln g\rangle$.

The validity of SPS in the theory of Abrahams et al. \cite{1} has generated debates and controversies in the past, which have
not been fully settled and which have recently been revived \cite{16,17,18}.  Indeed the justifications of the SPS hypothesis
in the analyses \cite{2,3,4,5} rests on the use of a random phase approximation (RPA) which assumes that the phases of the
amplitude reflection- and transmission coefficients $R_L$ and $T_L$ (with $r_L=|R_L|^2, t_L=|T_L|^2$) are uniformly distributed
over ($0,2\pi$)in the localization domain, which corresponds to length scales $L$ much larger than the localization length
$\xi$.  Although strong evidence, both numerical \cite{19} and analytical \cite{20,21}, exists for uniform phase distributions
for $L>>\xi$ in the 1D Anderson model \cite{22}, the controversy about the validity of SPS does not seem to be satisfactorily
resolved \cite{16,17,18,23}.  However, Cohen et al. \cite{23} have concluded that if RPA was not used one would obtain at most
two-parameter scaling.  Doubts about the RPA results have led Deych et al. \cite{16} to reconsider the scaling problem in the
framework of an exact solution for the Lloyd model, which differs from the Anderson model by the use of a Cauchy distribution
of site potentials.  While the study of the Lloyd model does not allow to draw conclusions for scaling in the Anderson model
(because the Cauchy distribution possesses infinite moments), the results of Ref. \cite{16} do show that the phase
randomness model is invalid for the Lloyd model.  On the other hand, Schomerus and Titov \cite{17} have studied the
deviations of the conductance distribution from the SPS form obtained in the RPA at the band centre ($E=0$).  They also
discussed \cite{17,18}, using large deviation statistics, the higher cumulants which, beyond the mean and the variance,
affect the exact distribution of conductance.

The purpose of the present paper is to calculate exactly the distribution of conductance in the localization regime
($L>>\xi$) in the current 1D Anderson model for weak disorder.  We feel that, in particular, our results definitely clarify the
status of the SPS hypothesis in the Anderson model.

The Schr\"{o}dinger equation  for a chain of $N$ disordered sites $1\leq m\leq N$ of spacing $a=1$ ($L=N$) is

\begin{equation}\label{eq2}
\varphi_{n+1}+\varphi_{n-1}+\varepsilon_n\varphi_n=E\;\varphi_n\quad ,
\end{equation}
where the site energies $\varepsilon_n$, in units of a constant hopping rate, are independent gaussian variables of zero mean
and correlation

\begin{equation}\label{eq3}
\langle\varepsilon_m\varepsilon_n\rangle=\eta^2 \delta_{m,n}\quad .
\end{equation}
The disordered chain is connected as usual at both ends to semi-infinite non-disordered chains ($\varepsilon_m=0$) with sites
$m>N$ and $m<1$, respectively.  

The distribution of the transmission coefficient $t_N$ (conductance) for an electron incident from the right with wavenumber
$-k$ (or energy $E=2 \cos k$) will be obtained by using the general recursion relations of Ref. \cite{21} which connect the
transmission- (reflection-) amplitudes of a chain of $n$ sites with the corresponding amplitudes for a chain with one less
disordered site of length $n-1$.  By iterating the transmission amplitude recursion relation (11.a) of \cite{21} we readily
find

\begin{equation}\label{eq4}
\ln T_N=i\;k\;N-\sum^N_{n=1}\ln[1-i\nu_n(1+e^{2ik}R_{n-1})]\quad,
\end{equation}
where the reflection amplitudes are given by Eq (11) of \cite{21}

\begin{equation}\label{eq5}
R_n=\frac{e^{2ik}R_{n-1}+i\nu_n(1+e^{2ik}R_{n-1}}{1-i\nu_n(1+e^{2ik}R_{n-1})}\quad.
\end{equation}
with

\begin{equation}\label{eq6}
\nu_n=\frac{\varepsilon_n}{2\sin k}
\end{equation}
The Eq. (\ref{eq4}) is our starting point for studying the cumulants of $\ln t_N=\ln T_N+\ln T_N^*$ and its probability
distribution.

For weak disorder (small $\varepsilon_m$) we restrict the analysis to low order in the correlation parameter $\eta^2$.  Since
the variance of $\ln t_N$ is found to be proportional to $\eta^4$ to lowest order an exact treatment for weak disorder must
involve expanding (\ref{eq4}) to 4th order in the site energies.  In parallel our analysis requires the reflection
amplitudes in (\ref{eq5}) to second order in the disorder, that is in the second Born approximation of backscattering.  Since
$R_m$ is linear in the site energies to lowest we expand it in terms of linear and quadratic contributions,
$R_m=R^{(1)}_m+R^{(2)}_m$, which are defined by recursion relations obtained by identifying contributions of the same order in
an expansion of both sides of (\ref{eq5}):

\begin{equation}\label{eq7}
R^{(1)}_n= e^{2ik}R^{(1)}_{n-1}+i\;\nu_n\quad ,
\end{equation}
\begin{equation}\label{eq8}
R^{(2)}_n= e^{2ik}R^{(2)}_{n-1}+2i\;\nu_ne^{2ik}R^{(1)}_{n-1}-\nu^2_n\quad .
\end{equation}
Solutions of (\ref{eq7}-\ref{eq8}) will be used for extracting explicit forms of successive order terms up to fourth
order in the expansion of $\ln t_N$ obtained from (\ref{eq4}),

\begin{equation}\label{eq9}
\ln T_N=\sum^N_{n=1}\sum^4_{p=1}(-1)^{p+1}\frac{u^p_n}{p}\;,\;u_n=-i\nu_n(1+e^{2ik}R_{n-1})\quad .
\end{equation}
At this point we have to determine the relevant asymptotic solutions of (\ref{eq7}-\ref{eq8}) for describing
transport in the strong localization regime, $N>>\xi$. In this domain the transport is dominated by the amplitudes $R_n$ at
large scales
$n$, which are described by some invariant distribution \cite{8}, which is independent of the initial values at sites where
the iteration of (\ref{eq5}) was started.  On the other hand, the probability densities associated with $R_n$ and $R_{n-1}$,
respectively, are related by a linear integral equation defined from the recursion relation (\ref{eq5}).  Since for large $n$
these densities are invariant it follows that in the recursion relation (\ref{eq5}) we may replace $R_{n-1}$ by $R_n$ in this
limit.  Then we obtain from (\ref{eq7}-\ref{eq8})

\begin{equation}\label{eq10}
R^{(1)}_n=\frac{i\nu_n}{1-e^{2ik}}\quad ,
\end{equation}
and
\begin{equation}\label{eq11}
R^{(2)}_n=\frac{\nu_n}{1-e^{2ik}}\left(\frac{2\nu_{n-1}}{1-e^{-2ik}}-\nu_n\right)\;,\; n>>\xi\quad .
\end{equation}
The existence of an invariant density for large $n$ is a general property which is equivalent to the ÒrandomÓ Oseledec theorem
\cite{24} (a review on these aspects and of their application in weak disorder studies is found in \cite{25}, see also
\cite{26}).

Using (\ref{eq9}-\ref{eq11}) we express the powers of $\ln t_N\;,\;(\ln t_N)^n\;,\;n=1,2,3,4$ to 4th order in the site
energies (powers higher than 4 have only contributions of higher order in the site energies) in order to find the moments,
using (\ref{eq3}).  In performing the averages we make use of the factorization property

\begin{equation}\label{eq12}
\langle\varepsilon_m\varepsilon_n\varepsilon_p\varepsilon_q\rangle
=
\langle\varepsilon_m\varepsilon_n\rangle
\langle\varepsilon_p\varepsilon_q\rangle
+
\langle\varepsilon_m\varepsilon_p\rangle
\langle\varepsilon_n\varepsilon_q\rangle
+
\langle\varepsilon_m\varepsilon_q\rangle
\langle\varepsilon_n\varepsilon_p\rangle\quad,
\end{equation}
valid for arbitrary indices $m,n,p,q$ different or not \cite{27}.  A useful simplification is also to note that
averages of the form $\langle\nu_nR^{(1,2)}_{n-1}\rangle$ vanish because $R^{(1,2)}_{n-1}$depends only on energies of sites
$m\leq n-1$.  By straightforward but somewhat cumbersome calculations we obtain the following final results for the moments:

\begin{equation}\label{eq13}
\langle\ln t_N\rangle=-2\gamma N\;,\;\gamma=\frac{\eta^2}{8\sin^2k}-\frac{\eta^4}{64\sin^4k}
\left(1-\frac{1}{2\sin^2k}\right)\quad ,
\end{equation}

\begin{equation}\label{eq14}
\langle(\ln t_N)^2\rangle=\frac{\eta^4}{16\sin^4 k}(N^2+3N)\;,\; N>>\xi\quad ,
\end{equation}

\begin{equation}\label{eq15}
\langle(\ln t_N)^n\rangle=O (\eta^6)\;,\; n\geq 3\quad ,
\end{equation}
which are exact within the second Born approximation.  The vanishing of the $n=3$ and $n=4$ moments (Eq.
(\ref{eq15})) through order $\eta^4$ results from the fact that the fourth order terms in both $(\ln t_N)^3$ and $(\ln t_N)^4$
cancel identically.  Finally from (\ref{eq12}) and (\ref{eq13}) we get for the second cumulant of $\ln t_N$:

\begin{equation}\label{eq16}
\text{var}\ln t_N=\frac{2\eta^4N}{16\sin^4 k}=12\gamma^2N+O (\eta^6)\quad ,
\end{equation}
and higher cumulants defined e.g. in \cite{27} vanish to fourth order, of course.

In discussing the above results we first note that $\gamma=\xi^{-1}$ in (\ref{eq13}) is nothing but the perturbation
expression for the inverse localization length, in which the leading term coincides with the well-known expression in the
(first) Born approximation \cite{28}.  Our results for the cumulants of $\ln t_L\simeq\ln g$ in the localized regime
$(N>>\xi)$ show that the distribution of the conductance has an exact log-normal form in the second Born approximation, namely

\begin{equation}\label{eq17}
P(g) dg=\frac{1}{\sqrt{24\pi\gamma^2N}}e^{-\frac{(\ln g+2\gamma N)^2}{24\gamma^2N}}d\ln g\quad .
\end{equation}
The important new feature of the exact result for the variance in (\ref{eq16}) is that while being
linear in $N$, in agreement with the central limit theorem, the ratio of it to the mean (\ref{eq13}) is proportional to the
inverse localization length.  In this case the criterion for SPS \cite{16}, which is obeyed in RPA, is clearly violated.  This
shows that the mean and the variance of $\ln g$ in the distribution (\ref{eq16}) act as distinct parameters even though they
obey the same scaling equation of the form $\frac{df}{d\ln g}=f\;,\;f=\langle\ln g\rangle$ or $\text{var}\ln g$.  In this
sense our exact analysis demonstrates two-parameter scaling in the localized regime of the Anderson model, in contrast to the
RPA which leads to a strictly SPS \cite{2,3,4,5}.

Our analytical results for the cumulants of $\ln g$ differ qualitatively from the results for the 1D Anderson model obtained
by Schomerus and Titov (ST) \cite{17}, who did not use RPA.  ST find that at the band center $(E=0)$ the first three cumulants
differ from zero and are proportional to $\langle\ln g\rangle$, which indicates that the distribution of $\ln g$ obeys a SPS. 
Furthermore, for large energies ST recover the SPS results of RPA \cite{17}.  We believe that the differences between
our results and those of \cite{17} originate in the fact that we are using an exact analysis of the discrete Anderson model
whereas the treatment of ST deals with a continuum limit of the tight-binding equations.  In particular, the disorder enters
in our analysis via the reduced site energies (\ref{eq6}) depending on the Bloch wavenumber $k$ at energy $E=2\cos k$.  On the
other hand, the length scale dependence of successive moments in (\ref{eq13}-\ref{eq15}) enters, quite distinctly of the
disorder parameters $\varepsilon_n$, via multiple summations over sites of the form (\ref{eq4}).  In contrast, in the
Fokker-Planck type continuum treatment of ST the disorder enters exclusively via a rescaled length and a reduced continuum
energy, which causes the cumulants at $E=0$ to depend on the rescaled length only (via $\langle\ln g\rangle$).

Finally, we note that the observed shortcoming of the RPA conductance (resistance) distribution in the localized regime
exists also in the well-known DMPK equation for the distribution of transmission eigenvalues in quasi 1D systems \cite{6,7}. 
This is because the DMPK equation reduces in the 1D limit to results obtained in the RPA.


\begin{thebibliography}{unsrt}
\bibitem{22} The invariant imbedding model studied in \cite{20} has recently been shown \cite{21} to be equivalent to the
continuum limit of the tight-binding Anderson model for weak disorder.
\end{thebibliography}
\end{document}